\documentclass[preprint,showpacs,preprintnumbers,amsmath,amssymb]{revtex4}

\usepackage{graphicx}
\usepackage{dcolumn}
\usepackage{bm}

\begin{document}

\preprint{Phys. Rev. Lett.  {\bf 96}, 045703 (2006)}

\title{Jahn-Teller Solitons, Structural Phase Transitions and Phase Separation}

\author{Dennis P. Clougherty}
\email{http://www.uvm.edu/physics/dpc/}
\affiliation{
Department of Physics\\
University of Vermont\\
Burlington, VT 05405-0125}

\date{November 1, 2005}

\begin{abstract}
It is demonstrated that under common conditions a molecular solid subject to Jahn-Teller interactions supports stable Q-ball-like non-topological solitons.  Such solitons represent a localized lump of excess electric charge in periodic motion accompanied by a time-dependent shape distortion of a set of adjacent molecules.  The motion of the distortion  can correspond to a true rotation or to a pseudo-rotation about the symmetric shape configuration.  These solitons are stable for Jahn-Teller coupling strengths below a critical value; however, as the Jahn-Teller coupling approaches this critical value, the size of the soliton diverges signaling an incipient  structural phase transition.  The soliton phase mimics features commonly attributed to phase separation in complex solids.
\end{abstract}


\maketitle


The inventory of electronic and optical materials with technological applications is continually expanding.  One theme that is emerging in these complex materials is the importance of the interplay between the electronic and lattice degrees of freedom in shaping the properties of the materials \cite{dagotto05}.  
While some kinds of electron-lattice interactions (deformation potential coupling, piezoelectric coupling, etc.) have been well-studied for decades, important qualitative differences can emerge in situations where there are electronic or vibrational degeneracies, a frequent occurrence in crystals with high spatial symmetry.  Jahn-Teller interactions involving the interaction between degenerate (or nearly degenerate) electronic and vibrational states are now thought to play a substantial role in many of the recently discovered materials systems, including high-temperature superconductors \cite{johnson1}, alkali-doped fullerides \cite{johnson2}, and the manganites \cite{dagotto}.

Here, a Jahn-Teller model is presented that supports non-topological solitons stabilized by a coupling of electron and lattice degrees of freedom.  The analogues of such states have been conjectured to have relevance to cosmological phase transitions in the early Universe \cite{gleiser}, and they have been cited as potential candidates for dark-matter \cite{darkmatter}.  Thus the materials discussed here are experimentally accessible systems where these states may be studied.  The soliton phase in this model provides an intrinsic mechanism for the kind of heterogeneous charge clumping \cite{bishop05} and ``phase separation'' found with increasing frequency in complex materials \cite{dagotto05}.

\paragraph*{The Model}
Consider a molecular crystal describing three electronic bands linearly coupled to a doubly degenerate distortion.  The electronic bands are constructed from an orbital doublet, transforming as $x$ and $y$ and a trivial orbital singlet; the doublet of distortions transforms as $x$ and $y$, and it is treated as a doublet of classical fields. The following continuum model Hamiltonian is taken 
\begin{equation}
H=H_e+H_{ph}+H_{JT}
\label{model}
\end{equation}
where
\begin{subequations}
\begin{equation}
H_e=\sum_{m=0,\pm 1} \int d^3x (-\frac{1}{2}\psi_m^\dagger \nabla^2 \psi_m+W\delta_{|m|, 1} \psi_m^\dagger  \psi_m)
\label{ee}
\end{equation}

\begin{equation}
H_{ph}=\sum_{m=\pm 1}\int d^3x (|\partial_t \phi_m |^2+\kappa_2 |\phi_m |^2+\kappa_3 |\phi_m |^3+\kappa_4 |\phi_m |^4)
\end{equation}

\begin{equation}
H_{JT}=-g\sum_{m=\pm 1} \int d^3x (\phi_{-m} \psi_m^\dagger  \psi_0 + H.c.)
\end{equation}
\end{subequations}
where $W$ is the energy splitting between the orbital doublet and the singlet, and $m$ labels the states in the axial angular momentum basis.  Atomic units ($\hbar=m_e=1$) are used throughout.

This is a continuum version of the pseudo-Jahn-Teller model with local symmetry $(A\oplus E)\otimes\epsilon$, and it is similar to a recently studied continuum model that was shown to contain new kinds of multiorbital polarons (``vector polarons'') \cite{dpc04} that feature  spatial variations of the vibronic mixing.  It is demonstrated here using a variational argument that the model of Eq.~\ref{model} admits non-topological Q-ball-like solitons \cite{coleman,lee} for a range of the model's parameters.

With a doubly degenerate vibrational spectrum, the model contains the following continuous symmetry 
\begin{subequations}
\begin{equation}
\phi_m\to \exp({i m \alpha})\phi_m
\end{equation}
\begin{equation}
\psi_{m}\to \exp({-i m \alpha}) \psi_{m}
\end{equation}
\label{sym}
\end{subequations}
where $m=0, \pm 1$.  This global symmetry
corresponds to conservation of axial angular momentum of the combined electron-lattice system.
The Noether charge associated with this global transformation is found to be given by
\begin{equation}
Q=\sum_{m}\int d^3x\ m (\psi_m^\dagger \psi_m +i \phi_m\partial_t \phi_{-m})
\label{noether}
\end{equation}

\paragraph*{Variational Ansatz}
Consider a trial form for the distortion fields with harmonic time-dependence 
\begin{equation}
\phi_{\pm 1}(\vec r,t)=\xi(\vec r) \exp(\pm i\omega t)
\end{equation}
corresponding to a spatially varying amplitude and a region with a local molecular distortion that rotates about the symmetric configuration.  Additionally the amplitude of the distortion $\xi(\vec r)$ is taken to be a spherical pulse in space, consistent with the Q-ball field amplitude in the thin wall approximation \cite{coleman}.  The amplitude and radius of the pulse are treated as variational parameters to be determined by minimizing the energy of the configuration subject to the constraint of fixed Noether charge.

In the frame co-rotating with the distortion, the Hamiltonian acquires an additional term responsible for rotation-electronic coupling, a term well-known in the effect of $\Lambda$-doubling \cite{landau}

\begin{equation}
H_r=-\omega \sum_m\int d^3x\ m \psi_m^\dagger \psi_m
\end{equation}
This additional term respects the symmetry transformations of Eq.~\ref{sym}, and the Noether charge $Q$ is also conserved in the corotating frame.

The Bogoliubov-de Gennes equations for the Hamiltonian system $H_e+H_{JT}+H_r$ are 
\begin{subequations}
\begin{equation}
 -\frac{1}{2}\nabla^2 u_0-g\xi (u_{-1}+u_{1})=\epsilon u_0
 \end{equation}
 \begin{equation}
-\frac{1}{2}\nabla^2 u_1+(W-\omega) u_1-g\xi u_{0}= \epsilon u_1
\end{equation}
\begin{equation}
-\frac{1}{2}\nabla^2 u_{-1}+(W+\omega) u_{-1}-g\xi u_{0}=\epsilon u_{-1}
\end{equation}
\label{bdg}
 \end{subequations}

In the case where $W \gg \omega$ and $\mu$ (the chemical potential), the system of equations reduces to 
\begin{subequations}
\begin{equation}
u_1\approx {g\xi \over W-\omega} u_{0}
\end{equation}
\begin{equation}
u_{-1}\approx {g\xi \over W+\omega} u_{0}
\end{equation}
\begin{equation}
 -\frac{1}{2}\nabla^2 u_0-{2g^2\xi^2 W\over W^2-\omega^2} u_0\approx \epsilon u_0
 \end{equation}
 \label{bdg2}
 \end{subequations}

\begin{figure}
\includegraphics[width=3in]{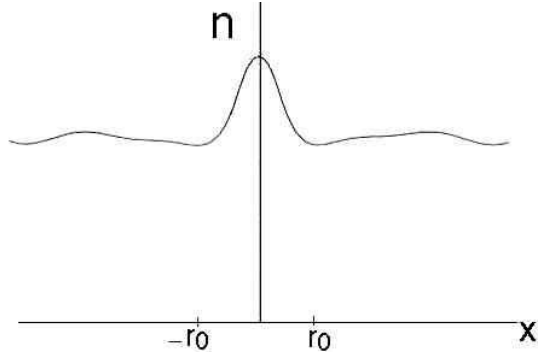}
\caption{\label{density} Electron charge density profile for a Jahn-Teller soliton.  Local distortions create an attractive potential well that enhances the electron charge density.  Friedel oscillations in the charge density set in beyond the distorted region $r > r_0$.}
\end{figure}

Electrons in the $m=0$ channel move in an effective square well potential that is attractive.  The strength of the potential varies inversely with $W$ and is proportional to the square of the distortion.  A Fermi gas experiencing this short-range attractive potential will form an electron-rich region in the vicinity of the potential with Friedel oscillations in the density that decay away with distance (see Fig.~\ref{density}).  

Within the Thomas-Fermi (TF) approximation for the kinetic energy, an analytic expression for the total kinetic energy of the electrons in the co-rotating frame can be found.  For $W$ large ($W\gg g \xi_0, \mu, \omega$), it is sufficient to only include the kinetic energy of electrons in the $m=0$ channel, and only terms $O((\omega/W)^2)$ need be retained.  
Dependence on $\omega$ can be eliminated in favor of the Noether charge $Q$, using the relation obtained from Eqs.~\ref{noether} and \ref{bdg2} 
\begin{equation}
Q\approx {4g^2\xi_0^2 \over W^3} n_0 \omega V.
\label{q}
\end{equation}
where $n_0$ is the electron density away from the distorted region and $V$ is the volume of the spherically symmetric distorted region.

Thus, 
\begin{equation}
E(V,\xi_0)=\frac{1}{2}(\kappa_2-{2g^2 n_0\over W})\xi_0^2 V+{3Q^2 W^3\over 16g^2n_0\xi_0^2 V}
+\kappa_3 \xi_0^3 V+\kappa_4 \xi_0^4 V
\label{energy}
\end{equation}
for the case where $\mu \gg g \xi_0$ where $\xi_0$ is the amplitude of the distortion \cite{note}.  

\paragraph*{Jahn-Teller Solitons}
The energy in Eq.~\ref{energy} is minimized with respect to the two variational parameters to give 
$V^2={\beta/\xi_0^2 U(\xi_0)}$ and $\xi_0=-{\kappa_3/ 2\kappa_4}$ where the effective potential $U(\xi_0)=\frac{1}{2}(\kappa_2-{2g^2 n_0/ W})\xi_0^2+\kappa_3 \xi_0^3 +\kappa_4 \xi_0^4$
and $\beta=3Q^2 W^3/16g^2n_0$.  Thus, a stable non-topological soliton is found for systems with negative cubic anharmonicity, a condition commonly invoked in models of the thermal expansion in solids.   The amplitude of displacement $\xi_0$ is determined only by elastic constants.  This is in contrast to other Jahn-Teller-induced displacive models \cite{bersuker,dpc00} where the amplitude of the displacement depends also on the strength of the electron-lattice coupling $g$.

Furthermore, it is noted that $U(\xi_0)$ vanishes at a critical coupling $g_c=\sqrt{W(2\kappa_2\kappa_4-\kappa_3^2)/ 4n_0\kappa_4}$ (see Fig. 2).  Thus the volume occupied by the soliton $V$ diverges when $g=g_c$.  As a consequence, it is seen from Eq.~\ref{q} that $\omega$ also vanishes at $g_c$ for finite charge, and the energy of the distorted state at $g_c$ becomes equal to the energy of the symmetric (undistorted) state.  Hence the system undergoes a structural phase transition for $g\ge g_c$.
It should be noted that for values of $g$ close to $g_c$, quantum fluctuations in the distortion become important and will alter the nature of this phase transition \cite{sachdev}.

\begin{figure}
\includegraphics[width=3in]{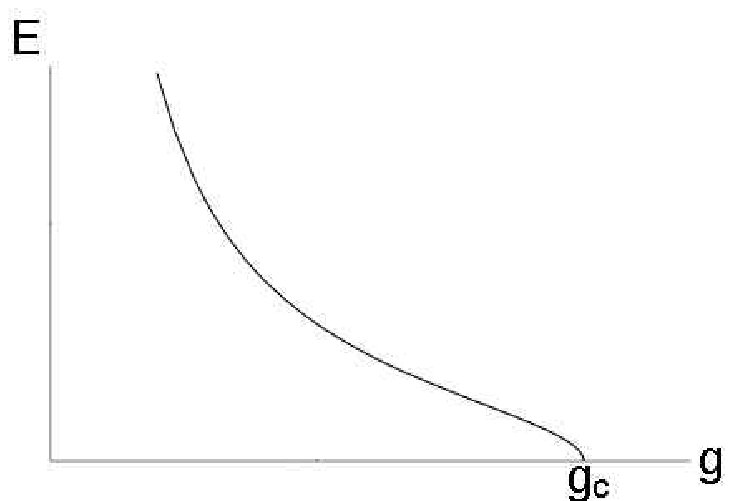} \includegraphics[width=3in]{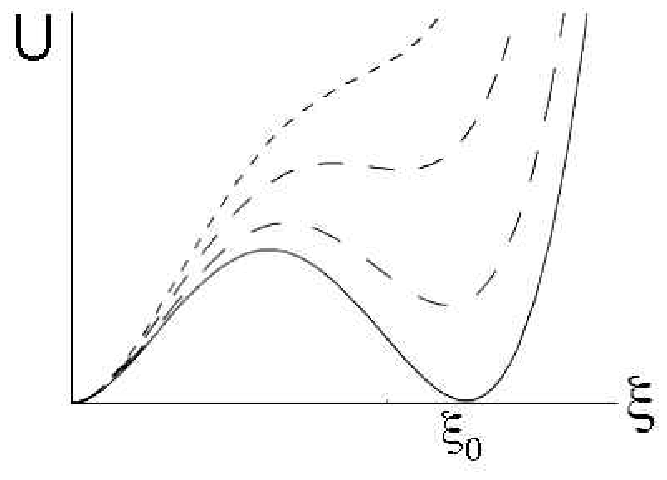}
\caption{\label{u} (a) Variational energy of Jahn-Teller soliton $E$ versus coupling constant $g$.  The soliton energy behaves as $\sqrt{g_c^2-g^2}/g$, thus vanishing as $g\to g_c$.  The vanishing soliton energy signals the onset of a structural phase transition. (b) Sketch of $U(\xi)$ for several values of $g$.  At $g_c$, $U$ vanishes at the local minimum at $\xi_0=|\kappa_3|/2\kappa_4$.}
\end{figure}

While only the volume and not the shape of the soliton alters the energy of this variational solution, the additional energy associated with a finite-sized wall, connecting the distorted region of the interior to the symmetric region exterior to the soliton, contributes a surface tension \cite{coleman}.   In the absence of vibrational or electronic anisotropy, the soliton is a sphere, the shape with minimum surface area.  

This model can be generalized in several ways.  The effects of electron-electron interactions might be included by the addition of a Hubbard-$U$ term in Eq.~\ref{ee}.  Within mean-field theory, the effects of the electron repulsion work counter to the lattice-induced attractive potential of Eq.~\ref{bdg2}; however, for a range of $U$ below a critical value $U_c$, the soliton will continue to be a stable solution of the model. 

Another generalization results from enlarging the symmetry group by coupling to larger multiplets.  Consider vibronically coupling a singlet band to a degenerate triplet of bands with a triplet of degenerate distortions, a continuum pseudo-Jahn-Teller model with local symmetry $(A\oplus T)\otimes\tau$.  The triplets support the vector representation of $SO(3)$.  Thus the vibronic interaction is invariant under the following continuous transformations:

\begin{subequations}
\begin{equation}
\phi_m\to \sum_{m'}{\cal D}^{(1)}_{m'm}(R)\phi_{m'}
\end{equation}
\begin{equation}
\psi_{m}\to \sum_{m'}{\cal D}^{(1)}_{m'm}(R^{-1}) \psi_{m'}
\end{equation}
\label{sym2}
\end{subequations}
where ${\cal D}^{(1)}_{m'm}(R)$ is the $j=1$ Wigner ${\cal D}$-matrix associated with the rotation $R$.  

For vibronic coupling below a critical value, the $SO(3)$ symmetry  is unbroken and a stable Jahn-Teller soliton results for a region of the parameter space.  The key ingredients in generalizing these results to other pseudo-Jahn-Teller models are adapted from Coleman \cite{coleman}: (1) the existence of an unbroken continuous symmetry, and (2) a global minimum in $U(\xi)/\xi^2$ for non-vanishing $\xi$.  For models with classical distortions, Jahn-Teller interactions (rather than pseudo) will not work even though many are invariant under continuous symmetries \cite{pooler}.  For such systems, the symmetry is broken for arbitrarily small coupling constant, thus violating the first criterion.

In summary, a continuum Jahn-Teller model was introduced that supports non-topological (Q-ball-like) solitons that self-localize through local electron-lattice interaction.  Such solitons are stable and are of finite size for coupling strength below a critical value.  The size of the soliton diverges as the coupling strength approaches this critical value, signaling a structural phase transition in the system.   It is anticipated that such states should be observable in two ways: (1) by measuring time-dependent fluctuations in atomic positions in molecular crystals that are close to a structural phase transition, and (2) by directly imaging the electron charge density.  A soliton phase in a molecular crystal would exhibit charge inhomogeneity and concomitant localized distortions, thus serving as a new kind of nanoscale structure relevant to a variety of complex materials.


\begin{references}
\bibitem{note} In general, part of the electron energy from the distorted region can be combined with the quartic contribution to the anharmonic phonon energy to renormalize  $\kappa_4$.  Here, this renormalization is ignored.
\end{references}
\end{document}